\documentclass[12pt]{article}
\pdfoutput=1
\usepackage{times}
\usepackage{epsfig}
\usepackage{amsmath}
\usepackage{amsfonts}
\usepackage{color}

\def\beq{\begin{equation}}
\def\eeq{\end{equation}}
\def\bea{\begin{eqnarray}}
\def\eea{\end{eqnarray}}
\def\beqn{\begin{eqnarray}} \def\eeqn{\end{eqnarray}}
\def\beeq{\begin{eqnarray}}
\def\eeeq{\end{eqnarray}}

\def\ep{\epsilon}

\def\nn{\nonumber}
\def\Eq#1{Eq.~(\ref{#1})}
\def\ln#1{\mathrm{log}\left(#1\right)}
\def\lnn#1{\mathrm{log^2}\left(#1\right)}

\def\bq{q\hspace{-.42em}/\hspace{-.07em}}

\def\td#1{\tilde{\delta}\left(#1\right)}

\newcommand{\la}{\langle}
\newcommand{\ra}{\rangle}

\def\pb{\mathbf{p}}

\def\M#1{{\cal M}^{(#1)}}

\def\uv{{\rm UV}}

\def\gs{g_{\rm S}}

\def\r{{\rm R}}

\def\ii{\imath 0}

\def\AA{{\cal A}^{(1)}}
\def\vev{\la v \ra}

\begin{document}

\begin{titlepage}
\renewcommand{\thefootnote}{\fnsymbol{footnote}}
\begin{flushright}
     IFIC/15-73 ,  TIF-UNIMI-2017-1
\end{flushright}
\par \vspace{10mm}
\begin{center}
{\LARGE \bf
Universal dual amplitudes and asymptotic expansions for $gg\to H$ and $H\to \gamma\gamma$  in four dimensions}
\end{center}
\par \vspace{2mm}
\begin{center}
{\bf 
F\'elix Driencourt-Mangin~$^{(a)}$\footnote{E-mail: felix.dm@ific.uv.es},
Germ\'an Rodrigo~$^{(a)}$\footnote{E-mail: german.rodrigo@csic.es}}
and {\bf
Germ\'an F. R. Sborlini~$^{(a,b)}$\footnote{E-mail: german.sborlini@unimi.it}}
\vspace{5mm}

${}^{a}$ Instituto de F\'{\i}sica Corpuscular, Universitat de Val\`{e}ncia -- Consejo Superior de Investigaciones Cient\'{\i}ficas, 
Parc Cient\'{\i}fic, E-46980 Paterna, Valencia, Spain. \\
${}^{b}$ Dipartimento di Fisica, Universit\`a di Milano and INFN Sezione di Milano, I-20133 Milano, Italy.
\end{center}

\par \vspace{2mm}
\begin{center} {\large \bf Abstract} \end{center}
\begin{quote}
Though the one-loop amplitudes of the Higgs boson to massless gauge bosons are finite because
there is no direct interaction at tree-level in the Standard Model, a well-defined regularization scheme is still required
for their correct evaluation. We reanalyze these amplitudes in the framework of the four-dimensional unsubtraction and the loop-tree duality (FDU/LTD), and show how a local renormalization solves potential regularization ambiguities.  
The Higgs boson interactions are also used to illustrate new additional advantages of this formalism. 
We show that LTD naturally leads to very compact integrand expressions in four space-time dimensions of the one-loop amplitude with virtual electroweak gauge bosons. They exhibit the same functional form as the amplitudes with top quarks and charged scalars, thus opening further possibilities for simplifications in higher-order computations.
Another outstanding application is the straightforward implementation of asymptotic expansions by using dual amplitudes. 
One of the main benefits of the LTD representation is that it is supported in a Euclidean space. 
This characteristic feature naturally leads to simpler asymptotic expansions.

\end{quote}

\par \vspace{5mm}

\vspace*{\fill}

\end{titlepage}

\setcounter{footnote}{0}
\renewcommand{\thefootnote}{\fnsymbol{footnote}}

\section{Introduction}
\label{sec:intro}
The $gg\to H$ and $H\to \gamma\gamma$ are the golden channels for production and decay 
of the Higgs boson at the CERN's Large Hadron Collider (LHC).
The one-loop contributions to the $Hgg$ vertex are known since a long
time ago~\cite{Wilczek:1977zn,Georgi:1977gs,Rizzo:1979mf},
as well as the Higgs decay into a photon pair~\cite{Ellis:1975ap,Ioffe:1976sd,Shifman:1979eb}. It is well-known that these amplitudes are finite
due to the absence of a direct interaction at tree-level in the Standard Model. 
However, contrary to what it is naively expected, dimensional regularization (DREG) 
-- or another regularization technique~\cite{renormworkshop} -- and 
a well-defined renormalization scheme are still required for their correct evaluation. Indeed, the naive calculation in four space-time dimensions not only leads to incorrect results~\cite{Gastmans:2011wh}, 
but also spoils gauge invariance and produces inconsistent physical effects such as the absence of 
decoupling in the limit $M_f^2 \gg M_H^2$, with $M_f$ the mass of the virtual particle running in the loop.

We have recently proposed a new approach to deal with perturbative computations
avoiding DREG. The four-dimensional unsubtraction method 
(FDU)~\cite{Hernandez-Pinto:2015ysa,Sborlini:2016gbr,Sborlini:2016hat} is based on an
integrand level transformation that achieves a fully local cancellation of singularities. 
The key component of this approach is the loop-tree duality theorem
(LTD)~\cite{Catani:2008xa,Bierenbaum:2010cy,Bierenbaum:2012th,Buchta:2014dfa,Buchta:2015wna},
which separates the loop contribution into sums of dual integrands obtained by applying a number of cuts equal to the number of loops. 

A generic one-loop amplitude with $N$-internal propagators has the form
\beq
{\cal A}^{(1)} = \int_\ell \left( \prod_{i=1}^N G_F(q_i)\right) \, {\cal N}(\ell, \{p_k\})~,
\eeq
where 
\beq
\int_\ell = -\imath \int \frac{d^d\ell}{(2\pi)^d} \, ,
\eeq
is the standard one-loop integration measure, $G_F(q_i)=(q_i^2-m_i^2+\ii)^{-1}$ are Feynman propagators, and ${\cal N}(\ell, \{p_k\})$ is the numerator that has a polynomial dependence in the loop and external momenta. The corresponding 
LTD representation is obtained by setting sequentially the internal propagators on-shell
\beq
{\cal A}^{(1)} = - \int_\ell \sum_{i=1}^{N} \td{q_i} \bigg( \prod_{j\ne i} G_D(q_i; q_j)\bigg) \, {\cal N}(\ell, \{p_k\})~,
\eeq
with $\td{q_i} = \imath \, 2\pi \, \theta(q_{i,0}) \, \delta(q_i^2-m_i^2)$, and promoting the 
remaining propagators to dual propagators with a modified $\ii$-prescription: 
$G_D(q_i; q_j) = (q_j^2-m_j^2 -\ii \, \eta\cdot (q_j-q_i))^{-1}$, where $\eta$ is a future-like vector, i.e. $\eta^2\ge 0$ and $\eta_0>0$.
From now on, we set $\eta_\mu = (1,\bf{0})$.

In this paper, we reanalyze the $g g \to H$ and $H\to \gamma\gamma$ scattering amplitudes at one-loop
and their asymptotic expansion in the FDU/LTD formalism. The four-dimensional nature of the FDU/LTD approach allows to get an alternative insight into the structure of these scattering amplitudes, unveiling the origin of local UV singularities that vanish in the integrated amplitude but lead to finite contributions. In the first place, we show how to apply the LTD theorem to obtain compact expressions for the amplitude integrand that exhibit the same functional form for virtual charged scalars, fermions (top quarks) or $W$ gauge bosons. This is a highly non-trivial result as intermediate expressions with 
gauge bosons diverge faster in the UV than those with scalars and fermions. 
After that, we discuss the local renormalization of the one-loop amplitude by 
introducing a suitable counter-term that locally cancels the UV behavior of the one-loop integrand and 
allows a direct integration of the amplitude in $d=4$ space-time dimensions. The second relevant result presented in this paper is related with asymptotic expansions. The simplicity and well-behaved convergence of the large-mass and small-mass asymptotic expansions of the Higgs boson amplitudes
in the LTD formalism avoids considering complementary expansions in different regions of the loop 
momentum~\cite{Beneke:1997zp,Smirnov:2002pj}.

\section{Dual amplitudes for $gg \to H$ and $H\to \gamma \gamma$}
\label{sec:dualamplitudes}
The one-loop scattering amplitudes of the Higgs boson to two massless gauge bosons have the form 
\bea
|\M{1}_{gg\to H} \ra &=&   \imath \, \gs^2 \, {\rm Tr} ({\bf T}^a {\bf T}^b) \, 
\varepsilon^\mu_a(p_1) \, \varepsilon^\nu_b(p_2) \, {\cal A}^{(1,t)}_{\mu\nu}~, \nn \\
|\M{1}_{H \to \gamma\gamma} \ra &=&   \imath \, e^2 \, \bigg( \sum_{f=\phi,t,W}  e_f^2 \, N_C^f \, {\cal A}^{(1,f)}_{\mu\nu} \bigg) 
\, (\varepsilon^{\mu}(p_1))^* \, (\varepsilon^{\nu}(p_2))^*~, 
\label{amplicolor}
\eea 
with $\varepsilon$ the polarization vectors of the external gluons and photons, 
${\rm Tr} ({\bf T}^a {\bf T}^b)= T_R \, \delta_{ab}$ the color factor, $N_C^f$ 
the number of colors, and $e_f$ the electric charge. Eventually, the sum in \Eq{amplicolor} might include 
the other quarks, and the leptons with $N_C^l=1$.
By Lorentz invariance, the color and electric charge stripped tensor amplitude is given by 
\beq
{\cal A}^{(1,f)}_{\mu\nu} =  \sum_{i=1}^{5} {\cal A}^{(1,f)}_i \, T_{\mu\nu}^i~,
\label{eq:scalarcoeff}
\eeq 
as a function of the tensor basis 
\beq
T^{\mu\nu}_i = \left\{ g^{\mu\nu} - \frac{2\,p_1^\nu \, p_2^\mu}{s_{12}}, g^{\mu\nu}~,  \frac{2\, p_1^\mu \, p_2^\nu}{s_{12}}~,
\frac{2\, p_1^\mu \, p_1^\nu}{s_{12}}~, \frac{2\, p_2^\mu \, p_2^\nu}{s_{12}} \right\}~,
\label{eq:tensors}
\eeq
where $s_{12}=(p_1+p_2)^2$, with $s_{12}=M_H^2$ if the Higgs boson is on-shell. We can extract the scalar coefficients ${\cal A}_i^{(1,f)}$ by using the projectors
\bea
P_1^{\mu\nu} &=& \frac{1}{d-2} \left( g^{\mu\nu} - \frac{2\, p_1^\nu p_2^\mu}{s_{12}}- (d-1)\, \frac{2\, p_1^\mu p_2^\nu}{s_{12}}\right)~,
\label{eq:projector}  \\
P_2^{\mu\nu} &=& \frac{2\, p_1^\mu p_2^\nu}{s_{12}}~,
\label{eq:projector2}
\eea
with $P_i^{\mu\nu} {\cal A}^{(1,f)}_{\mu\nu} =  {\cal A}_i^{(1,f)}$. 
Because of gauge invariance,
only the first coefficient ${\cal A}_1^{(1,f)}$ is relevant, while ${\cal A}_2^{(1,f)}$ should vanish 
upon integration. The other three coefficients do not contribute to the scattering amplitude
after contracting with the polarization vectors. 

The one-loop amplitude can be expressed in terms of the internal momenta 
$q_1 = \ell + p_1$, $q_2=\ell+p_{12}$ with $p_{12} = p_1+p_2$, $q_3 = \ell$; 
and $q_4=\ell + p_2$ to account for the diagrams with the two photons/gluons exchanged.
Explicitly, the one-loop amplitude with virtual top quarks is
\bea
{\cal A}^{(1,t)}_{\mu\nu} &=&  \frac{M_t}{\vev} \, \int_\ell \left(\prod_{i=1}^3 G_F(q_i) \right) \ \times
{\rm Tr} \left[  \gamma_\nu \, (\bq_1+M_t) \, \gamma_\mu \, (\bq_3+M_t)  \, (\bq_2+M_t) \right] + (p_1\leftrightarrow p_2)~, \
\label{original}
\eea
with $\vev$ the vacuum expectation value of the Higgs boson, and $G_F(q_i)=(q_i^2-M_t^2+\ii)^{-1}$ the Feynman propagators.

For the $W$ boson amplitude, we work in the unitary gauge because all
of the propagating degrees of freedom are physical and the internal propagators 
\beq
-\imath \left( g_{\mu\nu}- \frac{q_\mu q_\nu}{M_W^2} \right) \frac{1}{q_i^2-M_W^2+\ii}~,
\label{eq:propunitary}
\eeq 
do not introduce additional poles in the loop momentum space, 
allowing a straightforward application of the LTD theorem~\cite{Catani:2008xa}.
We do not provide here the explicit expressions equivalent to \Eq{original}
for the charged scalar and $W$ boson loop amplitudes; 
they can be obtained straightforwardly from the standard Feynman rules.

Partial results for the $W$ boson loop amplitude 
are more singular in the UV than the corresponding expressions for the charged scalar and top quark loops
due to the presence of higher powers of the loop momentum. 
These additional powers are introduced through the $WW\gamma$ vertex, which is linear in the loop momentum, 
and the $W$ propagator, as shown in \Eq{eq:propunitary}.
Also, the $W$ loop amplitude receives contributions from bubble diagrams with $WW\gamma\gamma$
interaction vertices that do not exist for the top quark loop. Those contributions 
are necessary to preserve gauge invariance. It is a remarkable feature of LTD that by setting the internal propagators 
on-shell the rank of the numerators of intermediate expressions is reduced automatically.
Moreover, the gauge invariance cross-cancellations between the bubble and triangle diagrams
explicitly arise in LTD without extra manipulations.

As a consequence, we obtain the following expressions for the scalar coefficients in \Eq{eq:scalarcoeff}
that exhibit the same functional form for charged scalars, fermions (top quarks) and $W$ bosons
\bea
{\cal A}_1^{(1,f)} &=& \, g_f \, \int_\ell \, \td{\ell} \bigg[
\bigg( \frac{\ell_{0}^{(+)}}{q_{1,0}^{(+)}} + \frac{\ell_{0}^{(+)}}{q_{4,0}^{(+)}} +  \frac{2 \, (2\ell\cdot p_{12})^2}{s_{12}^2-(2\ell\cdot p_{12}-\ii)^2} \bigg) 
\bigg( \frac{s_{12}\, M_f^2}{(2\, \ell \cdot p_1)(2\, \ell \cdot p_2)}\, c_1^{(f)} \nn \\ &+&  c_2^{(f)} \bigg) 
+ \frac{2 \, s_{12}^2}{s_{12}^2-(2\ell\cdot p_{12}-\ii)^2} \, c_3^{(f)} 
\label{universal1} 
\bigg]~, \\
{\cal A}_2^{(1,f)} &=& \, g_f  \, \frac{c_3^{(f)}}{2} \, \int_\ell \, \td{\ell} \, 
\bigg( \frac{\ell_{0}^{(+)}}{q_{1,0}^{(+)}} + \frac{\ell_{0}^{(+)}}{q_{4,0}^{(+)}} - 2 \bigg)~,
\label{universal2}
\eea
with $f=\phi, t, W$. The on-shell loop energies are given by
\bea
&& q_{1,0}^{(+)} = \sqrt{(\boldsymbol{\ell}+\pb_1)^2+M_f^2}~, \quad
q_{4,0}^{(+)} = \sqrt{(\boldsymbol{\ell}+\pb_2)^2+M_f^2}~,\nn \\ &&
\ell_0^{(+)} = q_{2,0}^{(+)} = q_{3,0}^{(+)} = \sqrt{\boldsymbol{\ell}^2+M_f^2}~.
\label{eq:eonshell}
\eea
It is worth mentioning that Eqs. (\ref{universal1}) and (\ref{universal2}) were derived from the application of the LTD theorem to the projected amplitudes in Eq. (\ref{original}) and the proper unification of the dual coordinate system, as carefully explained in Ref. \cite{Hernandez-Pinto:2015ysa,Sborlini:2016gbr}. The coefficients $c_i^{(f)}$  have indeed the form $c_i^{(f)} = c_{i,0}^{(f)} + r_f\, c_{i,1}^{(f)}$ with $r_f = s_{12}/M_f^2$. For the three different flavours that we consider, $f=\phi, t, W$, these coefficients are given by
\bea
&& g_f = \frac{2\, M_f^2}{\vev \, s_{12}}~,  \quad  c_{1,0}^{(f)} = \left( \frac{4}{d-2},-\frac{8}{d-2},\frac{4(d-1)}{d-2} \right) ~, \quad c_{1,1}^{(f)} = \left(0,1,\frac{2(5-2d)}{d-2} \right)~, \nn \\ 
&& c_{3,0}^{(f)} = (d-2) \, c_{1,0}^{(f)}~,  \quad c_{3,1}^{(f)} = 0~, \quad c_{23,0}^{(f)} = (d-4)\frac{c_{1,0}^{(f)}}{2}~, \quad c_{23,1}^{(f)} = \left(0,0, \frac{d-4}{d-2} \right)~,  
\eea
with  $c_2^{(f)}=c_{23}^{(f)}-c_3^{(f)}$. This result indicates that the calculation of the amplitude for other virtual states could
be reduced to the determination of the scalar coefficients $c_i^{(f)}$. 
The universality of the expressions in \Eq{universal1} and \Eq{universal2} could be supported by supersymmetric Ward identities at tree level similar to those relating
amplitudes with heavy quarks and heavy scalars~\cite{Schwinn:2006ca,Ferrario:2006np}, because 
the dual representation is indeed a tree-level like object. It is also interesting to notice that the two-loop amplitudes for scalar and pseudoscalar Higgs bosons 
to two photons have been calculated in Ref.~\cite{Harlander:2005rq} based on the assumption that 
if two physical processes correspond to a similar set of Feynman diagrams, then their cross sections
should be described by a common set of analytical functions. Their calculation is thus reduced 
to determine the coefficients of a linear combination of those functions  by solving a large set 
of linear equations arising from comparing the asymptotic expansions of a given ansatz and 
a one-dimensional integral representation of the amplitude. Such motivation could also be 
argued in this case, since similar physical processes should be described by similar integrand 
representations although with different coefficients. The LTD approach appears to be 
suitable for this purpose. We leave that discussion and the possible extension to two loops as an open question for a future publication.

Although the coefficient ${\cal A}_2^{(1,f)}$ vanishes upon integration in $d$-dimensions, notice that the naive calculation with $d=4$ leads to a finite contribution that 
violates gauge invariance. So, we can exploit that information to simplify the integrand-level expression for ${\cal A}_1^{(1,f)}$ by introducing non-trivial integrals which vanish in $d$-dimensions. Then, we can rewrite ${\cal A}_1^{(1,f)}$ in the most compact form: 
\bea
{\cal A}_1^{(1,f)} &=& \, g_f \, s_{12}\, \int_\ell \td{\ell} \bigg[
\bigg( \frac{\ell_{0}^{(+)}}{q_{1,0}^{(+)}} + \frac{\ell_{0}^{(+)}}{q_{4,0}^{(+)}} + \frac{2 \, (2\ell\cdot p_{12})^2}{s_{12}^2-(2\ell\cdot p_{12}-\ii)^2} \bigg) 
\frac{M_f^2}{(2\, \ell \cdot p_1)(2\, \ell \cdot p_2)}\, c_1^{(f)}  \nn \\
&+& \frac{2 \, s_{12}}{s_{12}^2-(2\ell\cdot p_{12}-\ii)^2}\, c_{23}^{(f)} \bigg]~,
\label{eq:compact}
\eea
which depends only on two independent coefficients. The integral proportional to the coefficient $c_1^{(f)}$ is indeed finite in the UV; thus, it will lead to the same result, up to ${\cal O}(\ep)$, if evaluated in four or $d$-dimensions. On the contrary, the remaining contribution must necessarily be calculated in $d$-dimensions because $c_{23}^{(f)} \propto d-4$ and the accompanying integral is logarithmically divergent. Since the coefficients $c_i^{(f)}$ depend on the nature of the particle circulating the loop through the associated Feynman rules, a consistent treatment of the dimensional extension of Dirac and Lorentz algebras is required to avoid potential mismatches in finite pieces. Explicitly, if we use the four-dimensional Dirac algebra by setting $d=4$ in the very first steps of the calculation, we get $c_{23}^{(f)}=0$ and the second line in Eq. (\ref{eq:compact}) is absent. However, a fully $d$-dimensional calculation shows that this terms leads to a finite non-vanishing contribution, arising from the UV pole of the bubble integral accompanying $c_{23}^{(f)}$. From the mathematical point of view, this behaviour is due to the absence of a local regularization, i.e. the integrand-level functions appearing in Eq. (\ref{eq:compact}) are not \emph{integrable}. In the next section, we will discuss how to implement a completely local renormalization to achieve integrability in four dimensions.

\section{Local renormalization and four-dimensional dual representation}
\label{sec:localrenormalization}
Since there is no direct interaction of the Higgs boson to massless gauge bosons, 
the one-loop amplitude is UV finite and does not need to be renormalized. However, 
the integrand of the one-loop amplitude is locally singular in the UV. This explains 
the requirement of introducing a well-defined regularization scheme to treat it properly. 
Therefore, we define in this section a UV counter-term that exactly cancels locally the 
UV behavior of the one-loop amplitude, but integrates to zero and does not lead to any 
effective renormalization. 

The approach that we follow in this paper differs slightly from the previously used in
Refs.~\cite{Hernandez-Pinto:2015ysa,Sborlini:2016gbr,Sborlini:2016hat}.
Instead of expanding around the UV propagator $G_F(q_\uv) = (q_\uv^2-\mu_{\uv}^2 + \ii)^{-1}$
at Feynman integral level, we first switch to the LTD representation and then expand. 
The UV expansion of \Eq{eq:compact} is particularly simple because only the contribution 
proportional to $c_{23}^{(f)}$ presents a singular behavior. The UV counter-term is defined as 
\beq
{\cal A}_{1, \uv}^{(1,f)} = -  g_f  \, s_{12} \, \int_{\boldsymbol{\ell}}  \frac{1}{4 (q_{\uv,0}^{(+)})^3}
\Bigg( 1 + \frac{1}{(q_{\uv,0}^{(+)})^2} \frac{3\, \mu_\uv^2}{d-4} \Bigg)\, c_{23}^{(f)}~,
\label{eq:A1dualgauge}
\eeq 
with $q_{\uv,0}^{(+)} = \sqrt{\boldsymbol{\ell}^2+\mu_\uv^2}$. The measure of the integral in \Eq{eq:A1dualgauge} is defined in the spatial components of the loop momentum, i.e. $\int_{\boldsymbol{\ell}}=\int d^{d-1}\boldsymbol{\ell}/(2\pi)^{d-1}$. The term proportional to $\mu_\uv^2$ is subleading in the UV limit and is used to fix the renormalization scheme. The factor $3/(d-4)$ has been adjusted to impose $\AA_{1,\uv} = 0$ 
in $d$-dimensions. This is the only place of the calculation where DREG is still necessary. Though, once the unintegrated UV counter-term has been computed, its four-dimensional limit can be used to regularize any other similar process. Notice that the renormalization scale $\mu_\uv$ is arbitrary because the one-loop amplitude 
is indeed not renormalized. For the scalar and the top quark amplitudes, the Dyson prescription~\cite{Dyson:1949ha}, which consists in
subtracting the amplitude evaluated with vanishing external photon (or gluon) momenta, has a similar effect. It fails, however, for the $W$ boson loop; 
although it correctly subtracts the leading  non-decoupling term in the limit $M_W^2/s_{12}\to \infty$, 
it does not account properly for the relevant subleading contributions. 

The difference of \Eq{eq:compact} and \Eq{eq:A1dualgauge} defines the locally renormalized amplitude ${\cal A}^{(1,f)}_{1,\r}$. Remarkably, it has a smooth four-dimensional limit and can directly be calculated with $d=4$,
\beq
\left. {\cal A}^{(1,f)}_{1,\r} \right|_{d=4} = \left( {\cal A}^{(1,f)}_1 - {\cal A}^{(1,f)}_{1,\uv}  \right)_{d=4}~. 
\label{eq:fourdimensions}
\eeq
Notice that $c_{23}^{(f)}$ vanishes in four dimensions and therefore the first term of the 
integrand in \Eq{eq:A1dualgauge} vanishes, but $\hat{c}_{23}^{(f)} = c_{23}^{(f)}/(d-4)$ leads to a finite contribution  
because $c_{23}^{(f)} \propto d-4$ but it is multiplied by an integral whose leading UV divergence behaves as $1/(d-4)$. 
Explicitly, the final and most compact expression for the unintegrated loop amplitude of this paper is
\bea
\nn \left. {\cal A}^{(1,f)}_{1,\r} \right|_{d=4}  &=& \, g_f \, s_{12} \, \int_{\boldsymbol{\ell}} \, \bigg[ \frac{1}{2 \ell_0^{(+)}}
\bigg( \frac{\ell_{0}^{(+)}}{q_{1,0}^{(+)}} + \frac{\ell_{0}^{(+)}}{q_{4,0}^{(+)}} + \frac{2 \, (2\ell\cdot p_{12})^2}{s_{12}^2-(2\ell\cdot p_{12}-\ii)^2} \bigg) 
\\ &\times& \frac{M_f^2}{(2\, \ell \cdot p_1)(2\, \ell \cdot p_2)}\, c_1^{(f)}  \ + \frac{3 \, \mu_\uv^2}{4 (q_{\uv,0}^{(+)})^5} \, \hat{c}_{23}^{(f)} \bigg]~,
\label{eq:compactandfourdimenional}
\eea
with the coefficients $c_{1}^{(f)}$ and $\hat{c}_{23}^{(f)}$ evaluated at $d=4$, i.e.
\beq
c_1^{(f)}= (2, - 4 + r_t, 6 - 3 \, r_W) \, , \ \ \ \ \hat{c}_{23}^{(f)}=(1, -2, 3+r_W/2) \, .
\eeq
The integrated amplitude reads
\beq
\left. {\cal A}^{(1,f)}_{1,\r} \right|_{d=4} = \frac{g_f\, s_{12}}{16\pi^2} \left( \frac{M_f^2}{s_{12}} \, \lnn{\frac{\beta_f-1}{\beta_f+1}} c_1^{(f)} 
+ 2 \, \hat{c}_{23}^{(f)}\right) ~, 
\eeq
with $\beta_f = \sqrt{1-4M_f^2/(s_{12}+\ii)}$, and it agrees with the expected well-known 
result~\cite{Wilczek:1977zn,Georgi:1977gs,Rizzo:1979mf,Ellis:1975ap,Ioffe:1976sd,Shifman:1979eb}. For the explicit integration, we have used the following parametrization of the loop three-momentum:
\beq
\boldsymbol{\ell} = M_f \, \xi \, (2 \sqrt{v(1-v)} \, {\bf e}_\perp, 1-2v)~,
\eeq
with $\xi$ its modulus normalized to the internal mass $M_f$, and ${\bf e}_\perp$ the unit vector in the transverse plane.
The dual integration measure is
\beq 
\int_\ell \td{\ell} =\int_{\boldsymbol{\ell}} \frac{1}{2 \ell_0^{(+)}} =  \frac{M_f^2}{4\pi^2}  \int_0^\infty  \xi_0^{-1} \, \xi^2 d\xi \, \int_0^1 \, dv~, 
\eeq 
with $\xi_0=\sqrt{\xi^2+1}$. The square roots present Eq. (\ref{eq:compactandfourdimenional}) can be transformed into rational functions of $x$ by implementing the change of variables
\beq
\xi = \frac{1}{2}\left(\sqrt{x}-\frac{1}{\sqrt{x}}\right) \, ,
\eeq
with $x\in [1,\infty)$.

\section{Asymptotic expansions in the Euclidean space of the loop three-momentum}
\label{sec:asymptotic}
LTD reduces the original $d$-dimensional integration domain with Minkowski metric, to a $(d-1)$-dimensional 
space with Euclidean metric: the loop momentum spatial components. In the particular case $d=4$, 
this Euclidean space corresponds to the domain of the loop three-momentum. This is an interesting feature that allows 
to circumvent potential difficulties that arise when performing asymptotic expansions of the integrand 
in a Minkowski space~\cite{Beneke:1997zp,Smirnov:2002pj}.
We can use the production and decay of the Higgs boson to massless gauge bosons as benchmark 
example to illustrate the ease of performing asymptotic expansions in the LTD formalism. The method is also applicable 
to other more complex processes. 

As starting example, we consider the large-mass limit $M_f^2 \gg s_{12}$ of the dual contribution with $q_3$ on-shell, i.e.
\beq
\td{q_3} \, G_D(q_3;q_2) = \frac{\td{q_3}}{s_{12} + 2q_3\cdot p_{12} -\ii}~,
\eeq
where $q_3\cdot p_{12}=q_{3,0}^{(+)}\sqrt{s_{12}}$ in the center-of-mass frame.
The on-shell energy $q_{3,0}^{(+)}$ is defined in \Eq{eq:eonshell}.
Since $q_{3,0}^{(+)} \ge M_f$, the asymptotic expansion for $M_f^2 \gg s_{12}$ is 
straightforwardly written as
\beq
\td{q_3} \, G_D(q_3;q_2) = \frac{\td{q_3}}{2q_3\cdot p_{12}}  \sum_{n=0}^\infty \left(\frac{-s_{12}}{2q_3\cdot p_{12}}\right)^n~.
\label{eq:exp1}
\eeq
Likewise, we shall expand the terms
\beq
\frac{\ell_0^{(+)}}{q_{1,0}^{(+)}} = \sum_{n=0}^\infty \frac{\Gamma(2n+1)}{\Gamma^2(n+1)} 
\left( -\frac{2\boldsymbol{\ell}\cdot \pb_1 + \pb_1^2}{(2 \ell_0^{(+)})^2}\right)^n~,
\label{eq:exp2}
\eeq
which is a valid expansion because $\ell_0^{(+)} > M_f$. Notice that each term of the 
expansions in \Eq{eq:exp1} and \Eq{eq:exp2} is less singular in the UV than the previous one, 
and is well-behaved in the IR. There is no need to consider additional loop momentum regions 
to obtain the correct asymptotic expansion. 

Taking into account the previous considerations, we obtain the following expansion 
 in the center-of-mass frame, again with $d=4$,
\bea
\left.  {\cal A}^{(1,f)}_{1,\r} (s_{12}<4M_f^2)\right|_{d=4} &=& \frac{M_f^2}{2\vev} \,  \int_{\boldsymbol{\ell}} \bigg[ 
\frac{M_f^2}{(\ell_{0}^{(+)})^5}
\bigg( \sum_{n=1}^\infty Q_n(z) \bigg( \frac{s_{12}}{(2\ell_0^{(+)})^2} \bigg)^{n-1} \bigg)   \, c_1^{(f)}
+ \frac{3 \, \mu_\uv^2}{(q_{\uv,0}^{(+)})^5} \, \hat{c}_{23}^{(f)}  \bigg]~, \nn \\
\label{asymptotic}
\eea
with $z=(2\boldsymbol{\ell}\cdot \pb_1)/(\ell_{0}^{(+)} \sqrt{s_{12}}) = \xi (1-2v)/\xi_0$, and 
\beq
Q_n (z) = \frac{1}{1-z^2} \left(P_{2n}(z) -1 \right)~,
\eeq
where $P_{2n} (z)$ is the Legendre polynomial. The asymptotic expansion of the amplitude 
in \Eq{asymptotic} can easily be integrated without using DREG. At the lowest orders, we find
\bea
\left.  {\cal A}^{(1,f)}_{1,\r} \right|_{d=4} &=& \frac{s_{12}}{8 \pi^2 \, \vev} \bigg( \frac{2 \hat{c}_{23,0}^{(f)}-c_{1,0}^{(f)}}{r_f} 
+ 2 \hat{c}_{23,1}^{(f)} - \frac{c_{1,0}^{(f)}}{12} - c_{1,1}^{(f)} 
- \bigg( \frac{c_{1,0}^{(f)}}{90} + \frac{c_{1,1}^{(f)}}{12} \bigg) \, r_f 
- \bigg( \frac{c_{1,0}^{(f)}}{560} + \frac{c_{1,1}^{(f)}}{90} \bigg) \, r_f^2 \nn \\ 
&-& \bigg( \frac{c_{1,0}^{(f)}}{3150} + \frac{c_{1,1}^{(f)}}{560} \bigg) \, r_f^3
- \bigg( \frac{c_{1,0}^{(f)}}{16632} + \frac{c_{1,1}^{(f)}}{3150} \bigg) \, r_f^4+ {\cal O} (r_f^5) \bigg)~.
\eea
The $1/r_f$ non-decoupling term vanishes because $\hat{c}_{23,0}^{(f)}=c_{1,0}^{(f)}/2$, 
leading to the following explicit results for the different internal particles: 
\bea
\left.  {\cal A}^{(1,\phi)}_{1,\r} \right|_{d=4} &=& \frac{s_{12}}{16 \pi^2 \, \vev} \bigg(-\frac{1}{3} - \frac{2}{45} \, r_\phi 
- \frac{1}{140} \, r_\phi^2 - \frac{2}{1575} \, r_\phi^3 - \frac{1}{4158} \, r_\phi^4+  {\cal O} (r_\phi^5) \bigg)~, \nn \\
\left.  {\cal A}^{(1,t)}_{1,\r} \right|_{d=4} &=& \frac{s_{12}}{16 \pi^2 \, \vev}  \bigg(-\frac{4}{3} - \frac{7}{90} \, r_t - \frac{1}{126} \, r_t^2  -  \frac{13}{12600} \, r_t^3  - \frac{8}{51975}\, r_t^4 +  {\cal O} (r_t^5) \bigg)~, \nn \\
\left.  {\cal A}^{(1,W)}_{1,\r} \right|_{d=4} &=& \frac{s_{12}}{16 \pi^2 \, \vev} \bigg(7 + \frac{11}{30} \, r_W 
+ \frac{19}{420} \, r_W^2 + \frac{29}{4200} \, r_W^3 + \frac{41}{34650} \, r_W^4 +  {\cal O} (r_W^5) \bigg)~. \
\eea

The asymptotic expansion for small $M_f$ can also be obtained from \Eq{eq:compactandfourdimenional} 
\bea
\left.  {\cal A}^{(1,f)}_{1,\r} (M_f^2\ll s_{12})\right|_{d=4} &=& \frac{M_f^2}{2\vev} \,  \int_{\boldsymbol{\ell}} \bigg[ 
\frac{-4 m_f^2}{(\ell_{0}^{(+)})^3 \, (1-z^2)} \bigg( 1 - 
\sum_{n=1}^\infty \frac{4 \, (\ell_{0}^{(+)})^2 \, \left(s_{12}\, m_f^2\, (2-m_f^2) \right)^{n-1}}
{\left(4 \, \boldsymbol{\ell}^2-s_{12}\, (1+m_f^2)^2\right)^n} \bigg) \, c_1^{(f)} \nn \\
&+& \frac{3 \, \mu_\uv^2}{(q_{\uv,0}^{(+)})^5} \, \hat{c}_{23}^{(f)}  \bigg]~, 
\label{asymptoticsmallM}
\eea
with $m_f^2=-M_f^2/(s_{12}+\ii)$. Once again, the terms of the expansion in \Eq{asymptoticsmallM} are less and less
singular in the UV at higher orders, allowing a full calculation with $d=4$. Integration of \Eq{asymptoticsmallM} leads to the awaited 
logarithmic contributions
\bea
\left.  {\cal A}^{(1,f)}_{1,\r} (M_f^2\ll s_{12})\right|_{d=4} &=& 
\frac{s_{12}}{8\pi^2 \, \vev}  \bigg( 2 \, \hat{c}_{23,1}^{(f)} - \left(2 \, \hat{c}_{23,0}^{(f)} + c_{1,1}^{(f)} \, L_f^2 \right) \, m_f^2
+ \left( c_{1,0}^{(f)} \, L_f^2  + 4 \, c_{1,1}^{(f)} \, L_f \right) \, m_f^4 \nn \\ 
&-& \left(4 \, c_{1,0}^{(f)} \, L_f + 2 \, c_{1,1}^{(f)}\, (2 + 3 \, L_f)  \right) \, m_f^6 \nn \\ 
&+& \left(2 \, c_{1,0}^{(f)}\, (2 + 3 \, L_f) + 4 \, c_{1,1}^{(f)} \left(3+\frac{10}{3} \, L_f  \right) \right) \, m_f^8 + {\cal O} \left(m_f^{10}\right) \bigg)~.  
\label{eq:small}
\eea
with $L_f = \ln{m_f^2}$.
As expected, the leading term in \Eq{eq:small} vanishes for charged scalars and top quarks 
since $\hat{c}_{23,1}^{(f)}=0$ for these particles,
but leads to a constant for the $W$ boson loop with $\hat{c}_{23,1}^{(W)}=1/2$.
Explicitly, 
\bea
\left.  {\cal A}^{(1,\phi)}_{1,\r} \right|_{d=4} &=& \frac{M_\phi^2}{8 \pi^2 \, \vev} \left( 2 
- 2  \, L_\phi^2 \, m_\phi^2+ 8 \, L_\phi \, m_\phi^4 - 4  \, (2+3 \, L_\phi) \, m_\phi^6 \right) +  {\cal O} (m_\phi^{10})~, 
\nn \\
\left.  {\cal A}^{(1,t)}_{1,\r} \right|_{d=4} &=& \frac{M_t^2}{8 \pi^2 \, \vev}  \bigg( - 4 + L_t^2
- 4  \, (1-L_t) \, L_t \, m_t^2 + 2 \, (2 - 5 \, L_t) \, m_t^4 \nn \\ &+& 4 \, \left(1+\frac{8}{3} \, L_t\right) \, m_t^6   \bigg) +  {\cal O} (m_t^{10})~, \nn \\
\left.  {\cal A}^{(1,W)}_{1,\r} \right|_{d=4} &=& \frac{s_{12}}{8 \pi^2 \, \vev} \big(1 - 3 \, (2-L_W^2) \, m_W^2  
- 6  \, (2-L_W) \, L_W \, m_W^4 + 6  \, (2-L_W) \, m_W^6 \nn \\ &-& 4 \, (3+L_W) \, m_W^8 + {\cal O} (m_W^{10}) \big)~, \
\eea
and these expressions are in agreement with the expansions shown in Ref. \cite{Aglietti:2006tp}.
In both cases -- the small and the large mass limits -- all the asymptotic expansions have been calculated directly in four space-time dimensions. 
This is achievable thanks to the fact that in the Euclidean space of the loop three-momentum 
it was necessary to consider a single kinematical region to achieve the correct asymptotic expansion in either of the two limits. In fact, this is a direct consequence of dealing with integrable and locally-regularized representations of the scattering amplitudes: there is an strict commutativity between integrals and parametric expansions at integrand-level.


\section{Conclusions}
\label{conclusions}
We have presented a very compact and universal integrand-level representation of the one-loop amplitude for the Higgs boson to two massless gauge bosons. The functional form of the amplitude is the same for internal scalars, fermions and vector bosons, and could be supported by tree-level supersymmetric Ward identities or be motivated from the fact that similar physical processes should be described by similar integrand representations with different coefficients. Presumably, this universality could be exploited further at higher orders.
The amplitude has been locally renormalized such that a pure four-dimensional expression free from potential scheme subtleties is obtained. All the previously known results were recovered within a pure four-dimensional representation of the loop amplitude. 

Since the integration of the FDU/LTD amplitude effectively occurs in an Euclidean space, namely the loop three-momentum space, asymptotic expansions are easily implemented. In fact, the local regularization in an Euclidean space implies that the series expansion of the integrand commutes with the integral symbol. Thus, expanding the integrand in any parameter (for instance, the mass of the particle circulating the loop) an integrating order-by-order, will lead to the right result. 
The asymptotic expansion of the Higgs boson amplitudes leads to very simple expressions that can easily be integrated. The results obtained, although focused on the Higgs boson interactions, can be generalized to other processes. In particular, the methods presented in this article open new possibilities for more efficient implementations and further simplifications of higher-order computations and asymptotic expansions.

\section*{Acknowledgements}

This work was supported by the Spanish Government and ERDF funds from the European Commission 
(Grants No. FPA2014-53631-C2-1-P and SEV-2014-0398), Generalitat Valenciana under Grant No. PROMETEO/2017/053, and
Consejo Superior de Investigaciones Cient\'{\i}ficas (Grant No. PIE-201750E021).
FDM acknowledges support from Generalitat Valenciana (GRISOLIA/2015/035) and GS from Fondazione Cariplo under the Grant No. 2015-0761.

\end{document}